\documentclass[twocolumn,prl,showpacs,preprintnumbers,amsmath,amssymb]{revtex4}
\usepackage[dvipdfm]{graphicx}
\usepackage{dcolumn}
\usepackage{bm}
\usepackage{amssymb}

\begin{document}
\title{Supersolid phases in the extended boson hubbard model}

\author{Kwai-Kong Ng}
\author{Y. C. Chen}
\affiliation{Department of Physics, Tunghai University, Taichung, Taiwan}
\date{\today}
\begin{abstract}
We present a comprehensive numerical study on the ground state
phase diagram of the two-dimensional hardcore boson extended
Hubbard model with nearest ($V_1$) and next nearest neighbor
($V_2$) repulsions. In addition to the quantum solid and
superfluid phases, we report the existence of striped supersolid
and three-quarter (quarter) filled supersolid at commensurate density
$\rho=0.75$ (0.25) due to the interplay of $V_1$ and $V_2$
interactions. The nature of three-quarter filled supersolid and
the associated quantum solid will be discussed.
 Quantum phase transition between the two supersolids of different
symmetries is observed and is clearly of first order.
\end{abstract}


\maketitle
Supersolid (SS) state \cite{Penrose,Andreev},
on which both diagonal and off-diagonal
 long-range order are broken, has been intensively
 discussed recently on
various models \cite{Batrouni,Sengupta,Wessel1,Ng,Laflorencie,Sengupta1}.
This is partially due to the experimental advance
of optical lattice that one day may investigate those hardcore
boson models and the exotic supersolid phase experimentally. Furthermore,
supersolid phase of spin models is also of great interests as
these quantum spin systems has been suggested could be
realized in real materials \cite{Ng,Laflorencie,Sengupta1}.

\begin{figure}
\includegraphics[width=6cm]{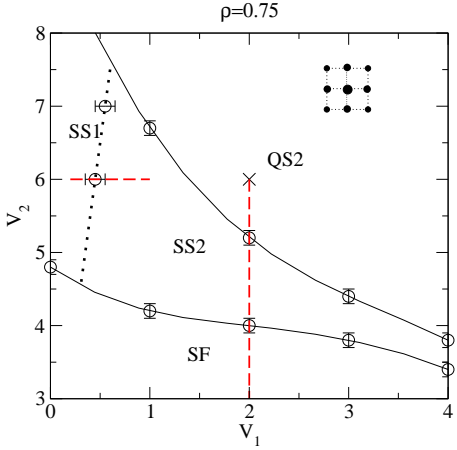}
\caption{(color online) Ground state phase diagram of $V_2$ as a function
of $V_1$ for density $\rho=0.75$. First (second) order phase transition
is denoted by dotted (solid) lines. The inset shows the boson
occupation profile (the star pattern) of the
QS2 and SS2 phase. The cross shows a representative
point of QS2 phase at $\rho=0.75$ where the order parameters are
plotted in Fig. \ref{fig2}.
Lattice size of 36x36 and 28x28 are used with temperature $\beta=1/2L$.}
\label{fig1}
\end{figure}

The simplest hardcore boson model that includes only the nearest
neighbor (nn) interaction, however, does not stabilize the
supersolid phase on square lattice \cite{Batrouni,Schmid}.
To induce the supersolid phase,
one may relax the hardcore constraint to softcore \cite{Sengupta}
or to include
the next nearest neighbor (nnn) interactions \cite{Batrouni}.
For the latter case,
a striped supersolid (SS1) phase is found associated with the
half-filled striped solid (QS1) phase, where bosons form
stripes that break the $x-y$ symmetry. This stripe structure
allows the superfluid component to easily flow through the
channels between stripes and therefore coexistence of both solid
and superfluid ordering is possible. Unlike striped solid,
the hardcore checkerboard solid provides no pathway for
superfluid component and no checkerboard SS of hardcore boson
has been found so far, unless nnn hopping is included \cite{Chen}.
In this work, we present a comprehensive study on the phase
diagram of hardcore boson hubbard model with nn and nnn interactions.
A three-quarter filled supersolid
that, like the checkerboard SS, preserves the $x-y$
symmetry is found stabilized in a large parameter regime of
$V_1$ and $V_2$. Associated to the supersolid is a
three-quarter filled quantum solid,
which share the same star-like occupation pattern
(see inset of Fig.\ref{fig1}).
For clarify, we call this solid and supersolid as the star
solid (QS2) and star supersolid (SS2) hereafter.
Interestingly, the supersolids, SS1 and SS2, that
possess different underlying symmetries compete
in some parameter regimes in which, as we will show,
first order phase transition occurs, in contrast to recent
work on a similar model \cite{Chen}. We tackle the problem with
both quantum Monte Carlo (QMC) and variational Monte Carlo (VMC)
methods which give consistent result. A generic Jastrow wave function
in VMC is able to generate qualitative features of all phases in
QMC calculations.

We study the extended boson Hubbard model on a 2D square lattice with the
Hamiltonian
\begin{equation}
H =  -t\sum_{i,j}^{nn} (b^\dagger_i b_j+b_i b^\dagger_j) + V_1\sum_{i,j}^{nn} n_i n_j + V_2\sum_{i,j} ^{nnn}n_i n_j - \mu \sum_{i} n_i
\end{equation}
where $b (b^\dagger)$ is the boson destruction (creation) operator
and $\sum^{nn} (\sum^{nnn})$ sums over the (next) nearest neighboring sites.
To set the energy scale of the problem, we let $t=1$ throughout
the paper.
At half filling, the ground state can be a checkerboard solid (with wave
vector ($\pi,\pi$)) for strong nn
coupling $V_1$, or a striped solid (with wave vector ($\pi,0)$ or (0,$\pi)$)
for strong nnn coupling $V_2$ \cite{Batrouni}. For
competing values of $V_1$ and $V_2$, however, quantum frustration disfavors
both solid structure and leads to the condensation of bosons
instead, i.e. a superfluid ground state.
Upon doping for large $V_2$, as mentioned above, striped solid structure
provides channels of superflow so that extra bosons
can form superfluid on top of the striped structure and
leads to a striped supersolid. Note that quantum effect
eventually drives all bosons to participate the superflow,
although the superfluidity transverses to the stripes
is much smaller \cite{Batrouni}.
For dominating $V_1$, on the other hand, addition bosons
forms no condensate on the checkerboard solid because
domain formation is energetically more favorable.
As a result, no supersolid of checkerboard solid ordering is found.
The phase diagram of half filling and result of doping close to
half filling have been discussed in detail in reference \cite{Batrouni}.

Remarkably, when further increasing doping to $\rho=0.75$,
our numerical calculations show a rich phase diagram that contains a
superfluid (SF) phase,
a star solid phase which have finite structure
factor $S(\textbf{Q})/N=\sum_{ij}
\langle n_i n_j e^{i \textbf{Q}\textbf{r}_{ij}}\rangle/N^2$
at $\textbf{Q}_0=(\pi,\pi)$, =$(\pi,0)$, and =$(0,\pi)$, and
supersolid phases of either star ordering or striped ordering.
The result obtained from QMC calculation with the stochastic series
expansion (SSE) algorithm \cite{Sandvik} on square lattice is
presented in Fig.\ref{fig1}. In QMC, the superfluidity, given by
$\rho_s=\langle W^2\rangle/4\beta t$ is computed by measuring the
winding number fluctuation.
The calculation is done on scanning over different $\mu$
grand canonically to search for the right $\mu$ that
fix the density $\rho=0.75$ for each coordinates ($V_1$,$V_2$) in the diagram.
Due to particle hole symmetry of the Hamiltonian $H$, one must obtain
the same phase diagram as in the Fig.\ref{fig1} for $\rho=0.25$.
Hereafter we will focus on $\rho=0.75$ but all discussion applies to
$\rho=0.25$ as well.
Let us now discuss each phases in more detail, starting with the
 star solid QS2.

\begin{figure}
\includegraphics[width=6.5cm]{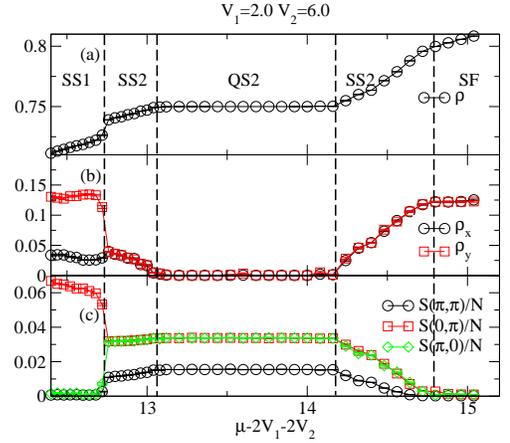}
\caption{(color online) QMC result of (a) boson density $\rho$,
(b) superfluidity $\rho_x$ and $\rho_y$, and
(c) structure factor of wavevectors
$(\pi,\pi)$, $(0,\pi)$ and $(\pi,0)$ as functions of chemical potential
$\mu-2V_1-2V_2$.
$V_1=2.0$, $V_2=6.0$ and lattice size is 28x28.}
\label{fig2}
\end{figure}

Inset of Fig.\ref{fig1} shows the ordering of the QS2. The lattice contains
four square sublattices with twice of the lattice constant.
It is important to stress that QS2 is not a
solid with three sublattices fully occupied and the fourth totally empty
which naturally give $\rho=0.75$.
Our calculation shows that all
sites have finite occupations as shown in the figure.
 As shown, two sublattices are identical
because of the $x-y$ symmetry. A typical structure of QS2 has one
of the sublattice almost fully occupied and the two identical sublattices have
occupation $n$ with the last sublattice has occupation $\sim2(1-n)$. At
$V_1=2.0$ and $V_2=6.0$, the occupations on different sublattices
are 0.99, 0.37, 0.37 and 0.27 respectively. It is not surprising that quantum
fluctuation and the gain in kinetic energy favor this ordering than the
ordering with one sublattice being empty.
One important feature
of the QS2 phase is that, although the $S(\textbf{Q}_0)/N$ is finite, it is
rather small compare to those of striped solid QS1. In Fig.\ref{fig2}, we show
the order parameters as a function of $\mu$ with $V_1=2$ and
$V_2=6$, a representative point (the cross in Fig.\ref{fig1})
of SS2 phase in the phase diagram, from which we obtain
all different phases by varying $\mu$.

A clear plateau appeared in the density curve signals
the existence of a solid phase, in which the superfluid density $\rho_{x(y)}$
 vanishes but $S(\textbf{Q}_0)/N$ (Fig.\ref{fig2})
is finite and remains flat throughout
the QS2 phase. Structure factor of all other wavevectors
are essentially zero but $S(\textbf{Q}_0)/N$ remains finite under finite
size analysis. It is note that the striped solid at half filling has
$S(\pi,0)/N \sim 0.2$ (not shown). The small value of $S(\textbf{Q}_0)/N$ in the
QS2, therefore, indicates the solid QS2 is rather \emph{soft}.

Away from the density $\rho=0.75$ there are supersolid phases
(SS2), as shown in Fig.\ref{fig2},
which is characterized by the same star solid
ordering (wavevector $\textbf{Q}_0$)
as the QS2 state and appears on both increasing or reducing
$\mu$ away from the QS2 phase. In this phase x-y symmetry is preserved
such that
$S(\pi,0)/N=S(0,\pi)/N$ and is about twice of $S(\pi,\pi)$.
All these peaks reduces
simultaneously away from the QS2 and vanish at the same
critical point where SF emerges at larger $\mu$.
This implies that
this SS2 is a unique phase characterized by these wavevectors but
not a mixture of striped phase. Remarkably, like the QS2 state,
SS2 has
one of the sublattices is almost fully occupied and does not
 participate on the superflow. Therefore
 the superfluid flows only on the other three sublattices.
 This self-pinning effect of one sublattice is rather rare
 and may worth further investigation \cite{Ng1}.
Transition from QS2 to SS2 is of second order as
both $\rho_{x(y)}$ and $S(\textbf{Q}_0)/N$ changes continuously
across phase boundaries and no abrupt change in order parameters
is observed.

On the other hand, there is clearly a first order phase transition
from SS2 to SS1 as shown in Fig.\ref{fig2}
where all parameters exhibit a sudden jump at
the $\mu-2V_1-2V_2\approx12.72$.
This discontinuous transition
arises from the distinct broken crystal symmetry of the
two supersolids.
 Our VMC
calculation, presented later,
also supports the discontinuous phase transition.
It is worthy noting
 that a recent study on the same model but with nnn hoping $t'$
 included also observes the QS2 and SS2 phases by Chen {\em et al.}
 \cite{Chen}.
 Our result indicates that $t'$ plays no significant role
 on the stabilization of both QS2 and SS2 phases which,
 indeed is a direct consequence of competition
 between $V_1$ and $V_2$ interactions. Furthermore, contrary
 to our findings, Chen {\em et al.} observe a crossover from SS1 to SS2.
 Whether it arises from $t'$ is still unclear yet.

\begin{figure}
\includegraphics[width=6cm]{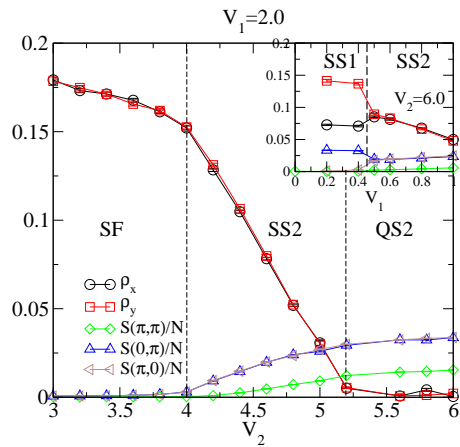}
\caption{(color online) Ground state order parameters
of a 28x28 lattice at $\rho=0.75$ for different $V_2$.
Inset shows the first order phase transition at $V_2=6.0$.}
\label{fig3}
\end{figure}

Moving down from the cross along the dotted
line in Fig. \ref{fig1}, the
width of the QS2 plateau shrinks as $V_2$ is reduced.
When $V_2\approx 5.2$
the QS2 phase disappears and the ground state changes
continuously to the SS2. Order parameters as functions of
$V_2$ are shown in the Fig. \ref{fig3} with $V_1=2$
(dotted line in Fig.\ref{fig1}). By reducing the nnn
repulsion, the system gains kinetic energy that favors
superfluidity and soften the solid structure at the same
time. Consequently, QS2 continuously changes to SS2 and to
SF eventually at $V_2=4.0$ where all peaks of $S(\bf{Q}_0)/N$
vanish simultaneously. Note that there is a large parameter
range where SS2 is stabilized at this commensurate density
$\rho=0.75$.

More complicated situation arises for $V_1 < 1$
where the nn repulsion is too weak to support the
star-like structure against the striped one. Fig. \ref{fig1}
shows the emergence of SS1 at small $V_1 < 1$ within
the SS2 regime. The phase transition between SS1 and SS2 is
again of first order because of the different broken translational
symmetries (inset of Fig.\ref{fig3}). For vanishing $V_1$,
neither SS2 and QS2 stabilized
so that only ground states of striped ordering is found for
all fillings, consistent to the previous findings \cite{Batrouni}.
In other words, the necessary condition for the appearance of
the star-like quantum solid and supersolid is the competition
between nn and nnn interactions.

\begin{figure}
\includegraphics[width=6cm]{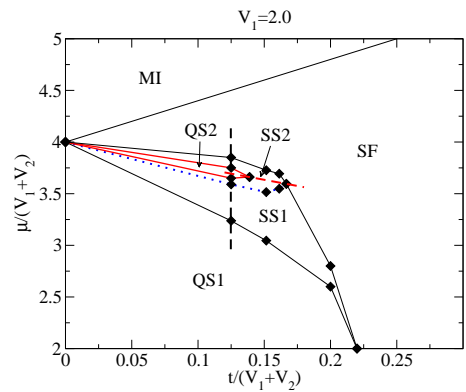}
\caption{(color online) Phase diagram of $\mu$
vs $t$ for fixed $V_1=2.0$. The lines are guide
to the eyes with dashed line (solid line) stands for
first (second) order phase transitions. The change
of order parameters along the black dotted line at $V_2=6.0$
is plotted in Fig. \ref{fig2} while those along
the red dotted at fixed $\rho=0.75$ is displayed in
Fig. \ref{fig3}.}
\label{fig4}
\end{figure}

To further investigate the effect of finite $V_1$, we plot in
Fig. \ref{fig4} the
phase diagram of fixed $V_1=2$ with varying $\mu$.
The phase diagram is similar to the case of vanishing $V_1$
(see ref.\cite{Batrouni})
except there are two new phases, SS2 and QS2, emerges within
the phase of SS1. Within this phase,
increasing $\mu$ (e.g. along the dashed line) such that $\rho$
approaches 3/4, the star-like ordering becomes
energetically more favorable than the striped ordering as
discussed before and SS2 or QS2 is stabilized.
Note that this happens
only when $V_2>4$, otherwise the SS1 phase simply
dissolves into SF phase upon increasing $\rho$.
The emergence of SS2, and QS2 in the phase diagram reflects
the interplay of nn and nnn interaction in the system.

\begin{figure}
\includegraphics[width=8cm]{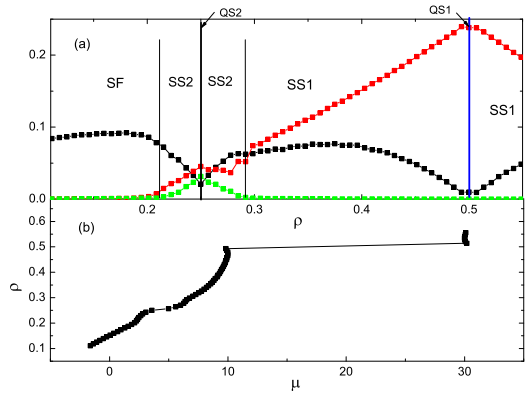}
\caption{(color online) VMC results of (a) the condensate
$N_{k=0}$(square), structure factor of wavevectors
$(\pi,\pi)$(circle) and $(0,\pi)$(triangle) as functions of
density and (b) boson density $\rho$ as a function of the chemical
potential. Here $V_1=4$, $V_2=6$ and lattice size is
$24\times24$.} \label{fig5}
\end{figure}

Now let us present our results by VMC in support of the star solid
QS2 and SS2 found in QMC calculation. The wave function we used is the
standard Jastrow wave function which is defined as :
\begin{equation}\label{wavefunction}
|\Psi \rangle = e^{-\sum_{i \neq j} v_{i,j} n_i n_j } |\Phi_0
\rangle,
\end{equation}
where $|\Phi_0 \rangle = (b_{k=0}^\dag)^N|0 \rangle$ is the
non-interacting superfluid wave function and $N$ is the total
number of bosons. In order to incorporate all kinds of phases in
the same wave function, the pair-wise potential $v_{i,j}$'s are
independently optimized by the algorithm proposed by Sorella
\cite{sorella}. To determine the phase diagram, we calculate the
number of bosons in the zero momentum mode, the condensate,
$N_{k=0}=b_{k=0}^{\dag}b_{k=0}$, $S(\pi,0)$ and $S(\pi,\pi)$ for
the optimized wave function. In Fig. \ref{fig5}(a) three order
parameters are shown as a function of density for given $V_1$=4
and $V_2$=6. As density increases, the phase changes from SF to
SS2 around the QS2 at $\rho$=0.25. With $\rho > 0.28$ $S(\pi,0)$
vanishes and the system becomes in SS1 phase and QS1 at $\rho$=0.5.
This is consistent with the phase diagram (Fig. \ref{fig1})
obtained by QMC. In order to verify if the new SS2 is
thermodynamically stable, we show the boson density as a function
of the $\mu$ which is calculated from the energy of
adding
add a particle to the system $\mu=E(N+1)-E(N)$. Two plateaux are
found at density $\rho$=0.25 and 0.5 which correspond to QS2 and QS1,
respectively. Positive slope around the plateau at $\rho$=0.25
manifests that the SS2 found is stable against phase separation.
Although there are discrepancies on the position of the phase
boundaries, the Jastrow wave function alone captures
the essential features
and successfully generates all the observed phases in QMC.

\begin{figure}
\includegraphics[width=8cm]{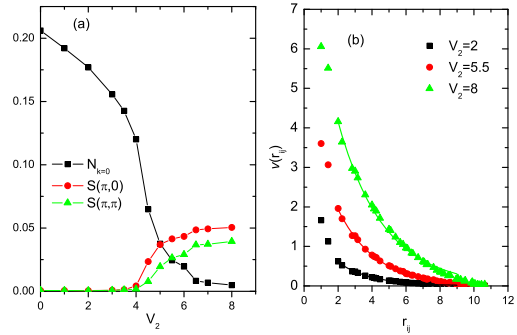}
\caption{(color online) (a) VMC results of the condensate
$N_{k=0}$, structure factor of wavevectors $(\pi,\pi)$ and
$(0,\pi)$ as functions of $V_2$ with $V_1=4$ and the boson density
n=0.25. The definition of the symbols are the same as Fig. 5. (b)
The variational parameters $v_{i,j}$ vs. $r_{ij}$ for $V_2$=2, 5.5
and 8. The solid lines are the fitting functions of $Ae^{-\xi r_{ij}}$.
The lattice size is $24\times24$.} \label{fig6}
\end{figure}

In Fig. \ref{fig6}(a) we show the order parameters as a function
of $V_2$ with $V_1$=4 and $\rho$=0.25. The
three phases found in Fig. \ref{fig3} for $V_2$=2 are also
observed here. The representative optimized variational parameters
$v(r_{ij})$ for SF($V_2$=2), SS2($V_2$=5.5) and QS2($V_2$=8) are
shown in Fig. \ref{fig6}(b). The data ($r\geq2$) is fitted to an exponential
form $Ae^{-r/\xi}$ with decay length $\xi$=2.0, 2.21 and 2.68 for SF, SS2 and QS2
respectively. The large value of $v(r_{ij})$ and $\xi$
in QS2 phase indicates the existence of a
strong long range repulsion between bosons while the
interaction is shorter ranged in SS2 and SF phases accordingly.

In summary, we present numerical evidences for the appearance
of a quantum solid, a supersolid phase with star pattern and
a striped supersolid
at or around $\rho=0.75$ or 0.25. The competition between
nn and nnn interaction is found to be important for observation
of both QS2 and SS2. A detail study
is given on the ground state phase diagrams by varying $V_1$ and $V_2$
as well as the chemical potential $\mu$.
The quantum phase transition between SS1 and SS2 is
appeared to be first order because of the abrupt change of
translational symmetry. Our VMC calculation also support the
QMC findings and the simple Jastrow wave function alone is adequate
to generate all the phases consistent with the QMC calculations.

\begin{acknowledgments}
The authors thank M.F. Yang for fruitful discussions and 
acknowledge
the support of the National Center for Theoretical Science.
K.K.N. acknowledges financial support by the NSC
(R.O.C.), grant no. NSC 95-2112-M-029-010-MY2 and NSC
95-2110-M-029-004.YCC is supported by NSC 95-2112-M-029-003-MY3.
Part of the calculation is supported by the National Center of
High Performance Calculation(Taiwan). 
\end{acknowledgments}

\end{document}